\documentclass{elsart}
\usepackage{natbib}
\begin{document}

\begin{frontmatter}
\title{A single population scheme for \\ FR\,I and FR\,II radio sources}
\author{Ignas Snellen \& Philip Best}
\address{Institute for Astronomy, University of Edinburgh, UK}

\begin{abstract}

The presence of two distant FR\,I radio galaxies
in the Hubble Deep Field plus Flanking Fields indicates that the 
number density of these objects is about 10$-$50 times higher at z$>$1
than in the local universe. This is in strong contrast with the 
idea that FR\,Is undergo no cosmological evolution. 
We advocate that the cosmological evolution of 
radio sources may be independent of FR class, and instead solely a function
of radio power. 
In this scenario the evolutionary properties of 
extragalactic radio sources can be fully described within a `single population
scheme'.
\end{abstract}


\end{frontmatter}
\section{Historic background}

Early statistics of radio sources indicated that the space density
of radio sources was much higher at early cosmological
epochs than in the present universe (Ryle \& Clarke 1961).
Subsequent deeper radio surveys showed a turnover in the number counts,
implying that this strong cosmological evolution must have been confined to
only the most powerful radio sources (Longair 1966), and a simple model
 was developed with two distinct population of radio sources; 
one low-luminosity non-evolving population, and one high-luminosity
strongly-evolving population (Doroshkevich, Longair \& Zeldovich 1970).
Wall (1980) suggested that these two populations of non-evolving and 
strongly evolving radio sources correspond to the 
Fanaroff \& Riley (1974) Class I and II galaxies respectively.
This idea of a `dual population scheme' has formed the basis of a 
comprehensive study by Jackson \& Wall (1999), who in addition linked
through orientation
the non-evolving and strongly evolving populations of FR\,I and FR\,II
radio galaxies to BL Lacs and flat spectrum quasars respectively.
A similar scheme has recently been developed by Willott et al. (2001), but 
here a weakly evolving and a strongly evolving population correspond to 
radio galaxies with low and high optical line luminosity, not to FR class. 

In the mean time, substantial differences were found between the host galaxies
of FR\,I and II radio sources. Owen \& Laing (1989) found that FR\,Is are
hosted by larger galaxies than FR\,IIs. In addition, Zirbel \& Baum 
(1995)
showed that at the same radio power, FR\,IIs have 5$-$30 times more 
optical line luminosity than FR\,Is. It was therefore proposed that
the FR dichotomy may be due to fundamental differences in the structural 
properties of the central engines in these two types of sources, such as
the accretion rate and/or the spin of the central black hole.

However, more
recently it has been shown by Ledlow and Owen (1996) that the
Owen \& Laing result
was caused by the combination of a sample selection effect and a
strong positive correlation between the FR\,I/II radio luminosity
cutoff and the absolute magnitude of the host-galaxy. This means that
FR\,I and FR\,II radio sources are actually found in very similar 
galactic environments.
In addition, Gopal-Krishna \& Wiita (2000) have pointed out a 
class of double radio sources in which the two
lobes exhibit clearly different FR morphologies. Although these objects 
are rare, their existence appears to be in conflict with the class 
of explanations that posit fundamental differences in the central engine.

\section{FR\,Is in the Hubble Deep Field}

If FR\,I and FR\,II radio galaxies undergo distinctly different  
cosmological evolutions, then observations and/or models which require a close 
link between the two classes of object are difficult to reconcile.
So why are the two FR classes thought to undergo different evolutions?
Firstly, the radio source counts undoubtly show that high luminosity 
sources undergo a much stronger evolution than sources of low 
radio luminosity. However, it is not at all clear whether this behaviour
is due to a gradual change in the evolution with radio
power, or caused by two distinct populations, one strongly evolving - 
assumed to be powerful FR\,IIs, and one non/weakly evolving - assumed 
in the Jackson \& Wall models to be FR\,Is.
Secondly, the V/V$_{\rm{max}}$ test is different for FR\,I and FR\,II radio 
galaxies, which has been brought forward as evidence for a difference
in evolution (eg. Jackson \& Wall 1999). However, in a flux density
limited sample such as 3C, low luminosity sources cover a much smaller
redshift range than high luminosity sources, and therefore 
the evolution signal is measured over a much larger range in 
cosmological epoch for FR\,II (z$<$2) than for FR\,I (z$<$0.2) radio 
galaxies. A fair comparison can only be made using the V/V$_{\rm{max}}$ test
if sources at similar radio power are compared. We show in 
Snellen \& Best (2001) that the V/V$_{\rm{max}}$ test differentiated
in radio power does not show any difference between FR\,Is and FR\,IIs
(see also Jackson \& Wall 2001), and therefore provides no evidence
that FR\,IIs evolve differently from FR\,I (at the same radio power).

It would be highly valuable if we would be able to directly measure the 
high redshift space density of FR\,I radio galaxies and 
subsequently their evolution. However, this is more easily said than done:
a luminous FR\,I radio galaxy at z=1 will only appear as a source
with a flux density of not more than a couple of milli-Janskys at 1.4 GHz. 
Moreover, to be able to 
morphology classify a radio source as an FR\,I at these redshifts
requires much deeper ($\sigma\approx10-20\mu$Jy) 
observations at (sub-)arcsecond resolution.

The only observations so far which comply to these limits are
those of the Hubble Deep Field (HDF) and Flanking Fields (HFF), which have
been imaged with the VLA at 8.4 GHz, with
the VLA and MERLIN at 1.4 GHz, with the 
WSRT at 1.4 GHz, and the EVN at 1.6 GHz
 (Richards et al. 1998; Muxlow et al. 2002; Garret et al 2000; 2001), 
all to $\mu$Jy flux density levels. 
Importantly, the two brightest radio sources in this 
field are distant FR\,I galaxies (Snellen \& Best 2001); one at z=1.101 
(log P$_{\rm{178MHz}}$=24.8 W Hz$^{-1}$sr$^{-1}$), and one 
without spectroscopic redshift but with such a faint and red magnitude that 
it must be at $z>1$(log P$_{\rm{178MHz}}>$25.3 W Hz$^{-1}$sr$^{-1}$).
We showed that it is very unlikely to find two
FR\,I radio galaxies at these luminosities and redshifts in such a small
area of sky if FR\,Is undergo no cosmological evolution. 
Consequently, the actual high redshift space density is estimated to be 
10$-$50 times higher than in the local universe (Snellen \& Best 2001).

\section{A single population scheme}

\begin{figure}
\includegraphics{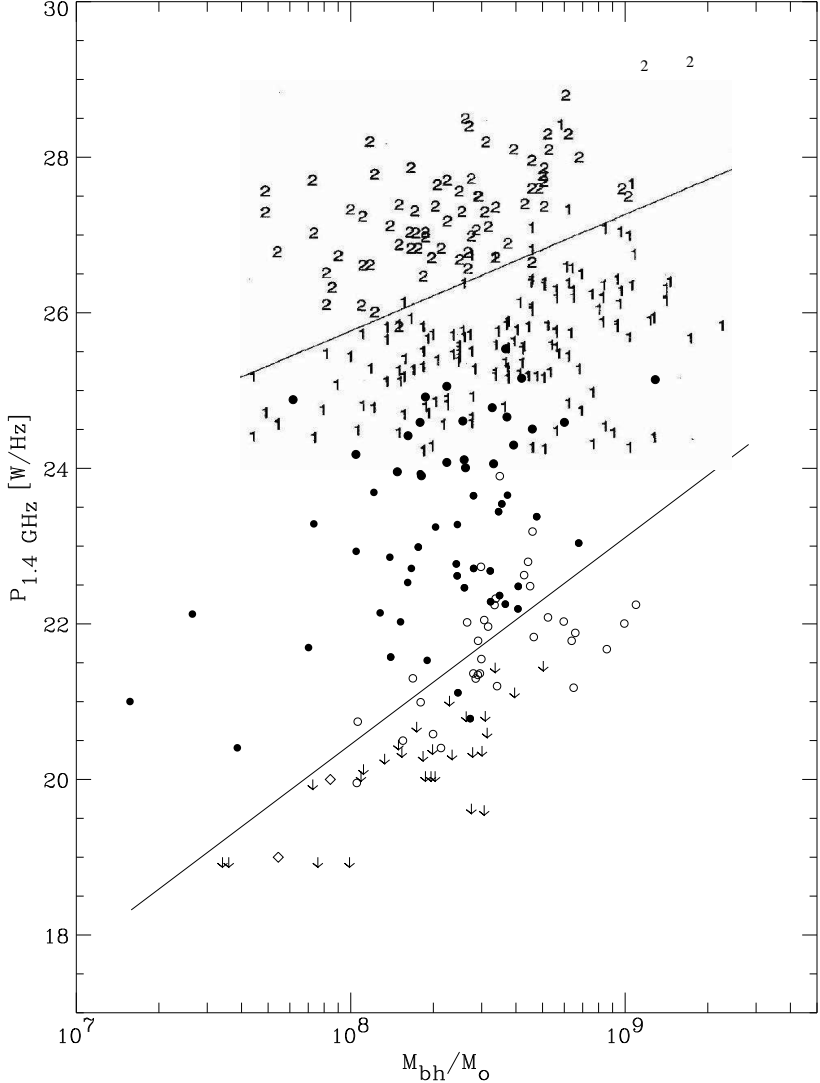}
\vspace{15cm}
\caption{\label{fig1} The 1.4 GHz radio power as function of 
black hole mass, for the FR\,Is (1) and FR\,IIs (2) of Ledlow \&
Owen (1996), and optically selected ellipticals from Faber et al. (1989),
 with the closed and open circles corresponding to those with 
extended and unresolved NVSS emission respectively. The upper line 
is the FR\,I/II cutoff as given by Ledlow \& Owen, and the lower 
line is the radio power - M$_{\rm{bh}}$ correlation as proposed by
Franceschini.}
\end{figure}

The results described above indicate that FR\,I radio galaxies 
{\it do} evolve with redshift.  We suggest that FR\,I and 
FR\,II radio galaxies should not be treated as intrinsically 
distinct classes 
of objects, but that the cosmological evolution is simply 
a function of radio power with  FR\,I and FR\,II radio galaxies of 
similar radio powers undergoing similar cosmological evolutions.
Since low power radio
galaxies have mainly FR\,I morphologies and high power radio galaxies
have mainly FR\,II morphologies, this results in a generally stronger
cosmological evolution for the FR\,IIs than the FR\,Is.

Basically, this ``single population scheme'' is similar to the
models used by Dunlop \& Peacock (1990), in which the 
evolution of the radio source population is only dependent on radio power.
Their models (a series of pure luminosity evolution and luminosity/density
evolution) fit the number counts and redshift distributions of bright
radio sources well. In addition they predict number densities which 
are comparable to that found in the HDF and HFF, and are consistent 
with the general mJy population in the LBDS Hercules sample, as shown
by Waddington et al. (2001).

The main benefit of treating the two FR classes as a single population
is that their physics can be closely linked.
A popular paradigm is that
the population of radio-loud AGN come with a range of jet outputs, 
of which the more powerful
may be strong enough to maintain their integrity  until they impact
on the intergalactic medium (IGM) in a shock. This results in an
FR\,II. However, if the jet is too weak, 
it may dissipate its energy by entraining IGM material, 
resulting in a more turbulent FR\,I. This may also explain the 
dependence of the FR\,I/FR\,II radio power divide on the luminosity
of the host galaxy: in more luminous galaxies, which
reside in denser environments or which have denser ISM, 
only jets of higher power can keep their integrity.
It also leaves open the possibility that during the lifetime
of a radio source, its morphology could change FR-class from
FR\,I to FR\,II or vice versa.

It may be interesting to note that the dependence of the 
FR cutoff on absolute magnitude of the host galaxy does not 
simply imply that this must be caused by an environmental effect. 
Since it has now been clearly established that the 
absolute magnitude of a galaxy is related to the 
mass of its central black hole (eg. McLure, this volume), 
the Ledlow \& Owen diagram also implies that 
FR\,I/II
cutoff depends on the mass of the central black hole; the 
more massive the black hole, the higher the FR\,I/II cutoff luminosity.
This is shown in Figure \ref{fig1}, where the radio luminosity is
plotted as function of black hole mass, M$_{\rm{bh}}$: 
for the radio sources from the Ledlow \& Owen (1996) sample M$_{\rm{bh}}$
is derived from the M$_{\rm{opt}}$-M$_{\rm{bh}}$ relation given in
Kormendy \& Gebhardt (2001), whilst for the optically selected
galaxies from Faber et al. (1989) M$_{\rm{bh}}$ 
is derived using the $\sigma$-M$_{\rm{bh}}$ relation found by
Gebhardt et al. (2000). Interestingly 
the slope of the FR\,I/II cutoff is very similar to that 
of the correlation between radio power and black hole mass as first 
proposed by Franceschini et al. (1996). The latter was initially 
believed to have only a small scatter, but subsequent work has shown that 
the scatter is at least several orders of magnitudes when 
powerful radio galaxies and quasars are included. 
It seems though that there is a lower limit to the radio power for a given 
black hole mass (eg. Dunlop et al. 2002; Snellen et al. 2002), and
maybe also an upper limit. This picture indicates the black hole mass or other 
fundamental properties of the central engine can not be ruled out of 
influencing the FR\,I/II divide.

\section{Future work: A new VLA survey}

We are continuing this field of research with a deep
wide-field survey conducted with the Very Large Array 
at 1.4 GHz in 
A configuration over two fields of 30$'$ diameter, using spectral line mode.
A noise level of  15 $\mu$Jy/beam is reach at a resolution of 1$''$.
In this way we sample a volume about an order of 
magnitude larger than that of the HDF+HFF survey. The fields
overlap with those of Windhorst et al. (1984; Lynx and Her.1), for
which previous low resolution radio observations and some
optical observations are available. To be able to morphologically
classify all radio sources, inevitably some objects will require to be 
followed up using the new VLA-Pietown link and/or MERLIN.

With this study we hope to firmly establish the cosmological space 
density evolution of FR\,Is, and investigate whether the evolution of 
radio-loud AGN is dependent on FR class, or solely a function of radio power. 
Secondly, we will determine possible variation of the FR\,I/II cutoff with 
redshift. This may shed new light on the physical connection between FR\,I 
and FR\,II radio galaxies.

\end{document}